\documentclass[
aip,jap,
reprint,
superscriptaddress,
amsmath,
amssymb
]{revtex4-1}

\usepackage{graphicx}
\usepackage{siunitx}
\usepackage{color}
\usepackage{tabularx}

 \definecolor{gray}{gray}{0.67}%helleres Grau
 \definecolor{Gray}{gray}{0.45}%dunkleres Grau
\begin{document}

\title{Band Offset in (Ga,In)As/Ga(As,Sb) Heterostructures}

\author{S.~\surname{Gies}}
\author{M.~J.~\surname{Weseloh}}
\author{C.~\surname{Fuchs}}
\author{W.~\surname{Stolz}}
\affiliation{Department of Physics and Material Sciences Center, Philipps-Universit\"{a}t Marburg, Renthof 5, 35032 Marburg, Germany}
\author{J.~\surname{Hader}}
\author{J.~V.~\surname{Moloney}}
\affiliation{Nonlinear Control Strategies Inc, 7040 N. Montecatina Dr., Tucson, AZ 85704, USA}
\affiliation{College of Optical Sciences, University of Arizona, Tucson, AZ 85721, USA}
\author{S.~W.~\surname{Koch}}
\author{W.~\surname{Heimbrodt}}
\email{wolfram.heimbrodt@physik.uni-marburg.de}
\affiliation{Department of Physics and Material Sciences Center, Philipps-Universit\"{a}t Marburg, Renthof 5, 35032 Marburg, Germany}

\date{\today}

%%%%%%%%%%%%%%%%%%%%%%%%%%%%%%%%%%%%%%%%%%%%%%%%%%%%%%%%%%%%%%%%%%%%%%%%%%%%%%%%%%%%%%%%%%%%%%%%%%%%%%%%%%%%%%%%%%%%%%%%%%%%%%%%%%%%%%%%%%%%%%%%%%%%%

\begin{abstract}

A series of (Ga,In)As/GaAs/Ga(As,Sb) multi-quantum well heterostructures is analyzed using temperature- and power-dependent photoluminescence (PL) spectroscopy. Pronounced PL variations with sample temperature are observed and analyzed using microscopic many-body theory and band structure calculations based on the \textbf{k$\cdot$p} method. This theory-experiment comparison reveals an unusual, temperature dependent variation of the band alignment between the (Ga,In)As and Ga(As,Sb) quantum wells. 

\end{abstract}

\maketitle

%%%%%%%%%%%%%%%%%%%%%%%%%%%%%%%%%%%%%%%%%%%%%%%%%%%%%%%%%%%%%%%%%%%%%%%%%%%%%%%%%%%%%%%%%%%%%%%%%%%%%%%%%%%%%%%%%%%%%%%%%%%%%%%%%%%%%%%%%%%%%%%%%%%%%

\section{Introduction}

The (Ga,In)As/GaAs/Ga(As,Sb) material system has recently been used for lasers operating over a wide spectral range in the infrared.\cite{Pan2010,Chang2014,Pan2013,Ripalda2005,Moeller2016}
Here, especially the W-type band structure has proven to be very effective as it allows for  more freedom in the device design\cite{Berger2015} and bears the promise for reduced Auger losses.\cite{Zegrya1995}
For optimized quantum design of the heterostructures with the goal to achieve good lasing operation, it is important to have profound knowledge of the band structure,\cite{Gies2015} in particular the band offsets.
For the \mbox{type-I}I systems under investigation, the band alignment can be determined with high precision if one uses both the spatially direct and the indirect transitions.\cite{Springer2016}
For this purpose, we present here a thorough photoluminescence (PL) analysis of the (Ga,In)As/Ga(As,Sb) \mbox{type-I}I heterostructures by means of temperature- and power-dependent PL spectroscopy.
Investigating different samples with a variation of the thickness of an internal GaAs barrier allows us to reveal the influence of tunneling processes that take place in these structures.
Analyzing the experimental results with our microscopic many-body theory, we are able to determine the temperature dependent band-alignment between the (Ga,In)As and Ga(As,Sb) wells. 
In the following Sec. II of this paper, we discuss the experimental details and results. The theoretical description and the PL calculations are presented in Sec. III, and a brief summary is given in Sec. IV, respectively.

\section{Experiment}

Our samples were grown in an AIXTRON AIX 200 GFR (Gas Foil Rotation) reactor system using metal-organic vapor-phase epitaxy (MOVPE).
The sample growth was carried out at a reactor pressure of \SI{50}{mbar} using H$_2$ as carrier gas and exact GaAs (001) substrates.
The growth process was carried out using triethylgallium (TEGa) and trimethylindium (TMIn) as group-III and tertiarybutylarsine (TBAs) and triethylantimony (TESb) as group-V precursors.
Before the layer structure was grown, a TBAs-stabilized bake-out procedure was applied in order to remove the native oxide layer from the substrate.
Afterwards, the reactor temperature was set to \SI{550}{ \celsius} for the growth of the active region as well as the barriers.
The former is composed of 5 times a multiple double quantum well heterostructure consisting of a \SI{5.2}{nm} thick (Ga$_x$,In$_{1-x}$)As and a \SI{5.0}{nm} thick Ga(As$_{1-y}$,Sb$_y$)  quantum well.
Both quantum wells are separated by a GaAs interlayer of variable thickness d. Each repetition of these double quantum wells is separated by a \SI{50}{nm} thick GaAs barrier.
High-resolution X-ray diffraction (HR-XRD, (004)-reflection) measurements were carried out in order to determine the structural properties of the samples.
The HR-XRD patterns were evaluated by fitting a full dynamical simulation to the experimental data.
The resulting compositions and interlayer thicknesses for each sample are given in table~\ref{tab:tab1}. 
Estimates of accuracy  of the chemical composition amounts to $\Delta$x$_{In}$ = $\pm$1.5\% and $\Delta$y$_{Sb}$ = $\pm$1.5\%, respectively.\\

The photoluminescence spectra have been aquired using a liquid nitrogen cooled Ge-detector and a \SI{0.5}{m} spectrometer, while the sample was excited using a frequency-doubled solid state laser at \SI{532}{nm}.
\begin{table}[h!t]%

\caption{Compositions and interlayer thicknesses of the investigated samples.}

\begin{center}

\begin{tabular}{ccc}

\hline

\hline

d (nm)& x$_{In}$ (\%) & y$_{Sb}$ (\%) \\

\hline

0		& 19.5 & 21.1 \\

0.4 & 20.7 & 21.1 \\

1.5 & 21.5 & 21.7 \\

3.5 & 21.0 & 23.8 \\

4.8 & 21.0 & 23.3 \\

\hline

\hline

\end{tabular}

\end{center}

\label{tab:tab1}

\end{table}

%%%%%%%%%%%%%%%%%%%%%%%%%%%%%%%%%%%%%%%%%%%%%%%%%%%%%%%%%%%%%%%%%%%%%%%%%%%%%%%%%%%%%%%%%%%%%%%%%%%%%%%%%%%%%%%%%%%%%%%%%%%%%%%%%%%%%%%%%%%%%%%%%%%%%

%\section{Results and Discussion}

\begin{figure}[!ht]

  \includegraphics[width=8.5cm]{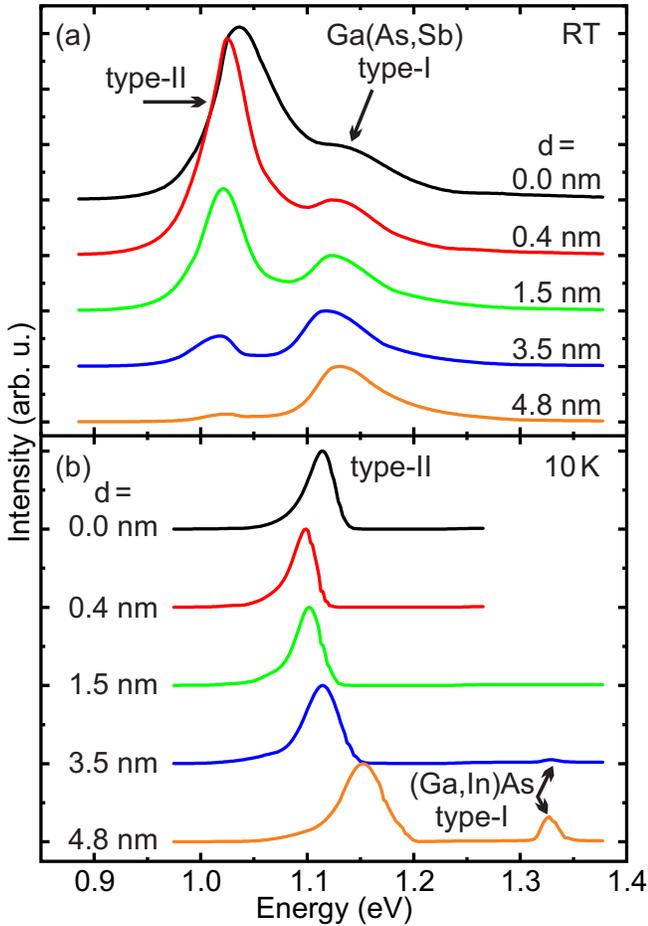}

  \caption{PL spectra of the five samples with different interlayer thicknesses.
	The RT PL spectra (a) are normalized to the Ga(As,Sb) peak at \SI{1.15}{eV} and shifted vertically, for clarity.
	At \SI{10}{K} (b) the spectra are normalized to the \mbox{type-I}I emission and also shifted vertically with respect to each other.}

  \label{fig:fig1}

\end{figure}

In Fig.~\ref{fig:fig1}(a) the room-temperature (RT) PL spectra are depicted for the five samples under investigation.
The spectra are normalized to the emission around \SI{1.15}{eV} attributed to the excitonic \mbox{type-I} recombination in Ga(As,Sb).\cite{Antypas1970,Nahory1977}

\begin{figure}

\includegraphics[width=8.5cm]{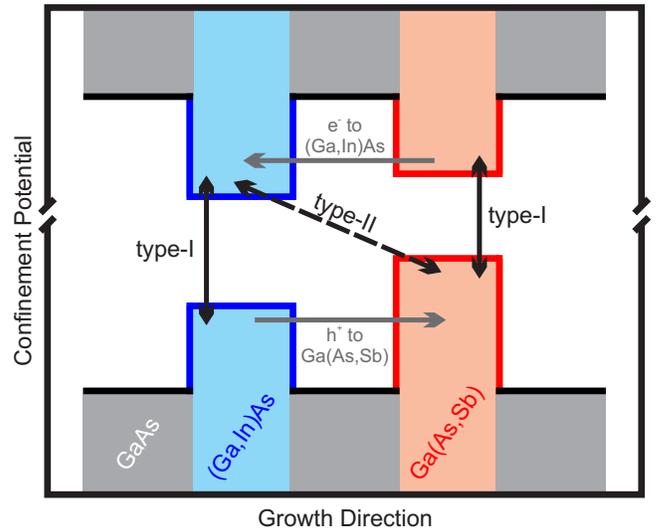}

\caption{Schematic depiction of the confinement potentials for electrons and holes in our samples.
The important recombination processes (black) are depicted as well as the tunneling directions of electrons and holes (gray).}%

\label{fig:fig2}

\end{figure}

At the low energy side another band at about \SI{1.05}{eV} can be seen. This band was found for all interlayer thicknesses but with varying intensity.
It is due to the \mbox{type-I}I charge-transfer  (CT) recombination of the electrons in the (Ga,In)As and the heavy-holes in the Ga(As,Sb).
For the sample without GaAs interlayer the \mbox{type-I}I transition exhibits the highest intensity, even higher than the \mbox{type-I} Ga(As,Sb) emission.
With increasing interlayer thickness d the \mbox{type-I}I PL decreases, as expected, due to the increasing spatial separation of electron and hole wavefunctions and the resulting decrease of the transition matrix element.  
At low temperatures the PL spectra shown in Fig.~\ref{fig:fig1}(b) differ strongly from the RT spectra.
The \mbox{type-I}I emission is the overwhelming band in the spectra, whereas the Ga(As,Sb) \mbox{type-I} PL does not occur.
For the samples with thickest interlayers of \SI{3.5}{nm} and \SI{4.8}{nm} an additional peak occurs at \SI{1.33}{eV}.
The origin of this transition is the (Ga,In)As \mbox{type-I} emission.\cite{Goetz1983}
To understand this behavior we need to take a look at the confinement potentials for electrons and holes in our samples, which are depicted schematically in Fig.~\ref{fig:fig2}.

\begin{figure}[!ht]

  \includegraphics[width=8.5cm]{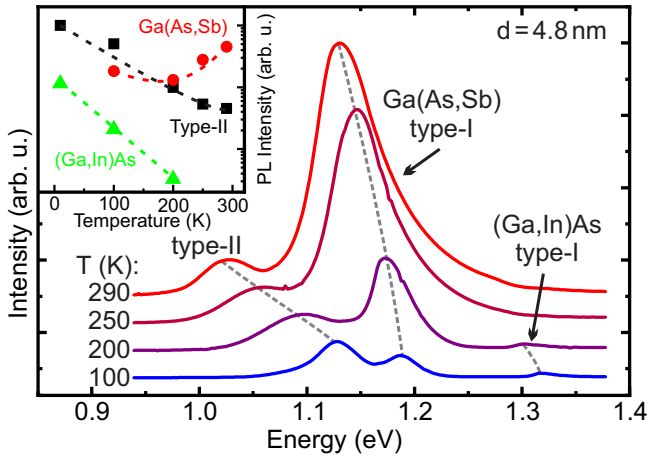}

  \caption{Temperature-dependent PL spectra of the sample with an interlayer thickness of $d = \SI{4.8}{nm}$.
	The spectra are normalized to the \mbox{type-I}I PL and shifted vertically, for clarity.
	In the inset the integrated intensity is shown of the \mbox{type-I}I, Ga(As,Sb) and (Ga,In)As PL as a function of temperature.}

  \label{fig:fig3}

\end{figure}

Besides the conduction band and valence band edges the relevant recombination processes (black) are given as well as the tunneling directions of electrons and holes (gray).
At \SI{10}{K} all charge carriers relax into the lowest energy states, i.e. the electrons are confined to the (Ga,In)As well and the holes are confined to the Ga(As,Sb) well (see Fig.~\ref{fig:fig2}).
The excitation energy was \SI{2.33}{eV}, that means well above the band gap of GaAs. After capturing of carriers by both wells, the subsequent relaxation process includes also a tunneling of electrons from the higher Ga(As,Sb) states into the (Ga,In)As QW. The holes may tunnel from the (Ga,In)As to the Ga(As,Sb) to reach the lowest states.
The recombination from the lowest states, respectively yield the \mbox{type-I}I PL that can clearly be seen in Fig.~\ref{fig:fig1}(b).
The occurrence of the (Ga,In)As \mbox{type-I} PL in the samples with thick interlayer at low temperature is due to the higher effective mass and the respective smaller hole tunneling probability from the (Ga,In)As to the Ga(As,Sb) compared to the electron tunneling in the opposite direction. 
There are obviously sufficiently many holes left to recombine radiatively in the (Ga,In)As well yielding a detectable PL.\cite{Tatebayashi2008}
Nevertheless, the (Ga,In)As PL is rather weak, especially compared to the \mbox{type-I}I transition, but gets stronger for thicker internal barriers where the tunneling probability is further decreased.
In opposite, the Ga(As,Sb) \mbox{type-I} PL cannot be observed in the cw-experiments as the electron tunneling probability is much higher than that for the holes.
To further study this behavior we did temperature dependent PL measurements for the sample with d = \SI{4.8}{nm}. The results are depicted in Fig.~\ref{fig:fig3}.
The PL spectra are normalized to the \mbox{type-I}I emission and shifted vertically for clarity.

It can be seen clearly, that the \mbox{type-I}I CT-PL is visible at all temperatures but undergoes a redshift of about \SI{120}{meV} with increasing temperature between \SI{10}{K} and \SI{290}{K}.  
Its integrated intensity decreases with increasing ambient temperature as can be seen in the inset of Fig.~\ref{fig:fig3} (black squares). This is expected behavior due to enhanced radiationless losses caused by an increasing phonon coupling. 
A very similar behavior was found for the (Ga,In)As \mbox{type-I} luminescence. At low temperatures it can be observed around \SI{1.33}{eV}.
With increasing temperature this peak also shifts red and decreases in intensity (green triangles in the inset).
At \SI{200}{K} the emission becomes so weak that it cannot be detected anymore, either due to the increasing exciton-phonon interaction at elevated temperatures or due to a phonon enhanced tunneling probability of the holes. 
The Ga(As,Sb) \mbox{type-I} emission, however, exhibits a totally different behavior. It can not be detected at 10 K as can be seen in Fig.~\ref{fig:fig1}(b) but is observable clearly in the spectrum at \SI{100}{K} and increases further with increasing temperature. This appearance can be understood considering that the Ga(As,Sb) electron states get populated thermally at higher temperatures.
We come back to this point in section III.

Here, it is interesting to note, that the redshift of the \mbox{type-I}I CT-PL (\SI{100}{meV}) between \SI{100}{K} and \SI{290}{K} is much bigger than the shift of the Ga(As,Sb) \mbox{type-I} PL, which shifts only about \SI{60}{meV}.
This reveals, that upon increasing the temperature the CB and VB edges in the (Ga,In)As and Ga(As,Sb) shift differently with respect to each other.
This results eventually in a temperature dependent offset between (Ga,In)As and Ga(As,Sb), as further discussed in section III.

\begin{figure}[!ht]

  \includegraphics[width=8.5cm]{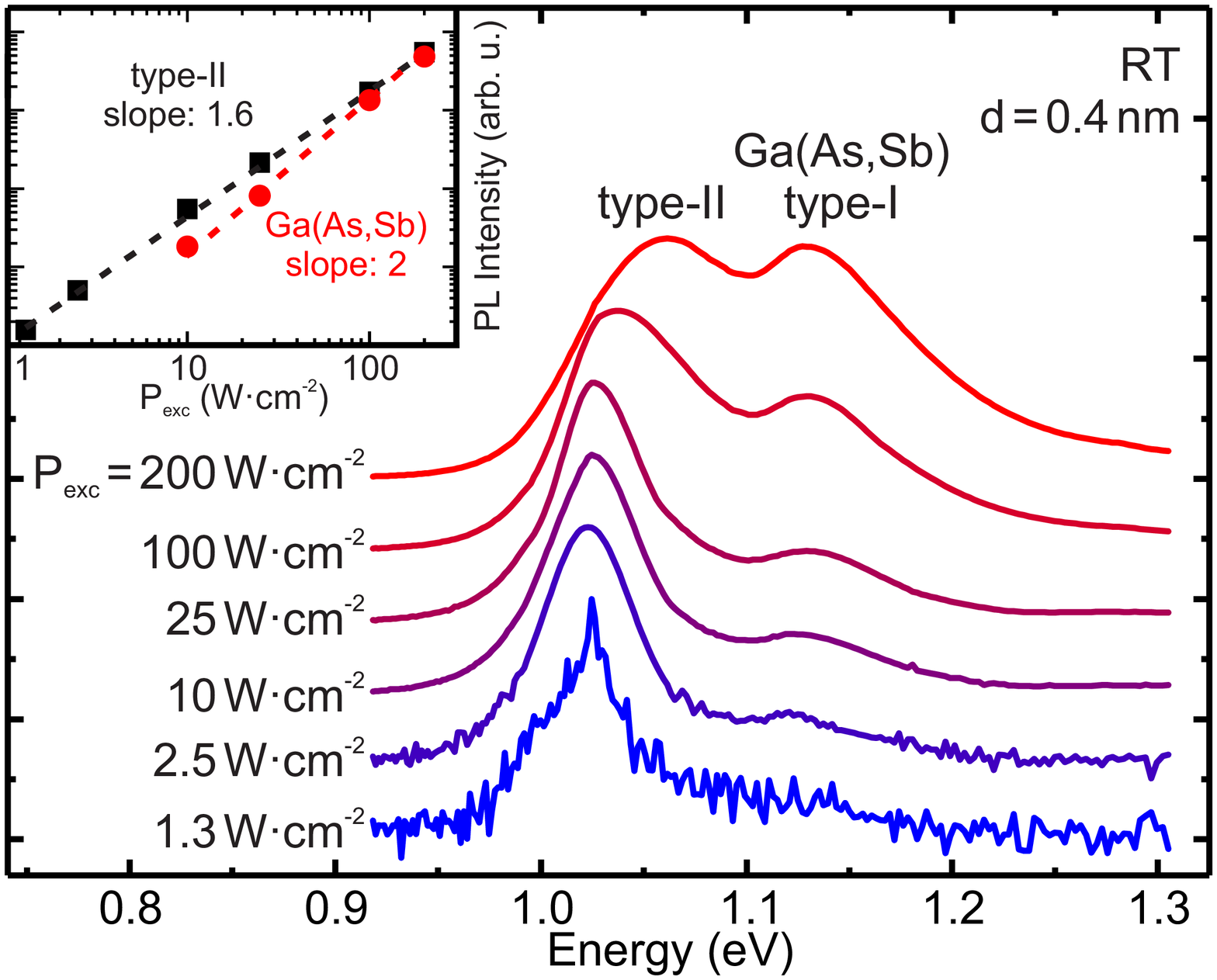}

  \caption{PL spectra of the sample with $d = \SI{0.4}{nm}$ for different excitation power between \SI[inter-unit-product = \ensuremath{{}\cdot{}}]{1.3}{\watt\per\square\centi\metre} and  \SI[inter-unit-product = \ensuremath{{}\cdot{}}]{200}{\watt\per\square\centi\metre}.
	The spectra are normalized to the \mbox{type-I}I emission band and shifted vertically with respect to each other.
	In the inset the PL intensity of the \mbox{type-I}I (black) and the Ga(As,Sb) (red) emission is depicted in a log-log plot.
	}
	
  \label{fig:fig4}
  
\end{figure}

From the temperature dependence it gets obvious, that the thermal carrier distribution has a strong impact on the PL properties. To further investigate the correlation between the carrier distribution and the \mbox{type-I}I PL, we studied the excitation power dependency of the (Ga,In)As/GaAs/Ga(As,Sb) material system.
In Fig.~\ref{fig:fig4} the PL spectra are depicted of the sample with $d = \SI{0.4}{nm}$ for different excitation powers between \SI[inter-unit-product = \ensuremath{{}\cdot{}}]{1.3}{\watt\per\square\centi\metre} and  \SI[inter-unit-product = \ensuremath{{}\cdot{}}]{200}{\watt\per\square\centi\metre}.

The PL spectra for the various excitation densities are normalized to the \mbox{type-I}I emission and shifted vertically for clarity.
Both, the \mbox{type-I}I and the Ga(As,Sb) emission undergo a blueshift with increasing excitation power due to band filling. The slightly stronger shift of the \mbox{type-I}I PL is caused by the band bending effect due to the carrier accumulation\cite{Springer2016}. From the inset in Fig.~\ref{fig:fig4} it can be seen, that the slope in the log-log plot of PL intensity as a function of excitation density for the Ga(As,Sb) \mbox{type-I} PL is 2, while the slope for the \mbox{type-I}I emission (about 1.6) seems to be slightly smaller. 
We come back to the excitation density behavior in the following chapter.

%%%%%%%%%%%%%%%%%%%%%%%%%%%%%%%%%%%%%%%%%%%%%%%%%%%%%%%%%%%%%%%%%%%%%%%%%%%%%%%%%%%%%%%%%%%%%%%%%%%%%%%%%%%%%%%%%%%%%%%%%%%%%%%%%%%%%%%%%%%%%%%%%%%%%

\section{Theory}

%Intro u Berechnungsgrundlagen

The theoretical PL is calculated using the semiconductor luminescence equations (SLE)\cite{ Kira1999189}. Here, the used Coulomb matrix elements, the dipole matrix elements, and the single particle energies are calculated using $8\times 8$ \textbf{k$\cdot$p} theory\cite{PhysRevB.55.6960,ChowKoch}.
The electron-electron and electron-phonon scattering is taken into account at the level of the second-Born-Markov approximation \cite{Hader2003513}.\\
The band gaps are calculated using parameters from \cite{vurg2001}.
In order to achieve good agreement with the experimentally measured PL of the samples with interlayer thicknesses of 
d=\SI{4.8}{nm} and d=\SI{3.5}{nm}, the experimentally determined well compositions were varied within the limits of the uncertainities from the HR-XRD analysis. 
For the numerical simulations of the samples with d=\SI{4.8}{nm}, the Sb concentration was set to \SI{23.7}{\%}  while the In concentration was set to \SI{20.3}{\%}, respectively. 
Simulating the sample with d=\SI{3.5}{nm}, concentrations of \SI{23.6}{\%} Sb and \SI{22.4}{\%} In were used. For both samples, it is assumed that the Ga(As,Sb) conduction band edge is situated energetically lower
than that of GaAs, see Fig. \ref{fig:fig2}. 
Structural disorder effects are taken into account by inhomogeneously broadening the PL spectra by convolution with a Gaussian distribution with 
full width at half maximum of \SI{30}{meV}. 

\begin{figure}

 \includegraphics[width=8.5cm]{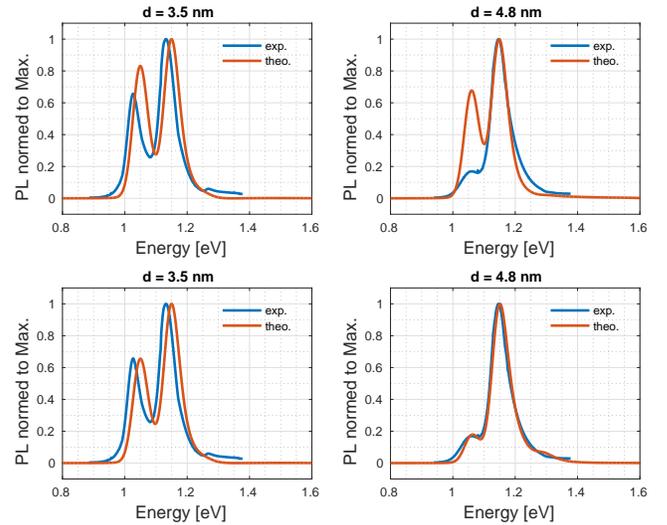}

 \caption{Experimental and theoretical PL of the samples with d=\SI{3.5}{nm} (left) and d=\SI{4.8}{nm} (right).
 	  In the upper part, the carrier temperature matches the lattice temperature of \SI{250}{K}.  
	  In the lower part the carrier temperature for the theoretical curve is higher, \SI{265}{K} for the sample with d=\SI{3.5}{nm} (left)  
	  and  \SI{380}{K} for the sample with d=\SI{4.8}{nm} (right).}

 \label{fig:Peakheights}

\end{figure}

In order to achieve a satisfactory agreement with the experimental results, the different band offset were ad- justed.
Considering the sample with d=\SI{4.8}{nm}, at \SI{250}{K} a Ga(As,Sb) conduction band offset (CBO) of \SI{5.5}{\%} of the total band offset relative to GaAs and a (Ga,In)As CBO of \SI{64.83}{\%} 
reproduces the experimentally measured \mbox{type-I}I peak positions, see Fig. \ref{fig:Peakheights}. 

\begin{figure}

 \includegraphics[width=8.5cm]{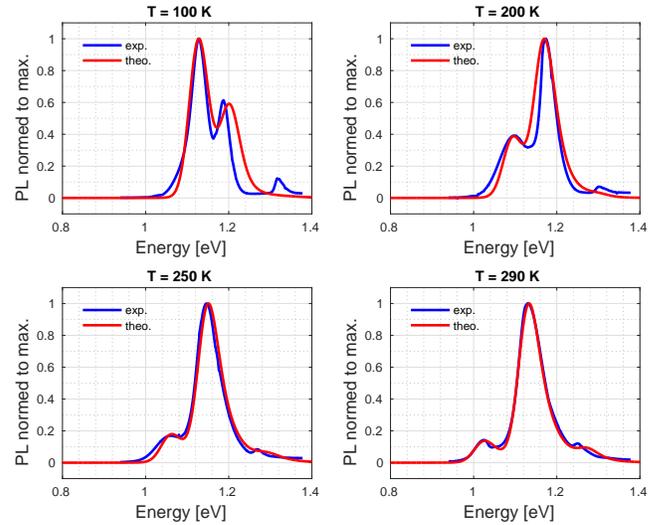}

 \caption{ Experimental and theoretical PL of the sample with d=\SI{4.8}{nm}. 
  The temperatures mentioned in the titles are the lattice temperatures. The carrier temperature is increased compared to the corresponding lattice temperature. 
  }

 \label{fig:theoPL_26662}

\end{figure}

>From the theory-experiment comparison concerning the temperature dependence (see Fig.~\ref{fig:theoPL_26662}), we deduce that the conduction band edge of the (Ga,In)As well and the 
valence band edge of the Ga(As,Sb) well shift in that way, that they increase their energetic separation with decreasing temperature. This was realized by a variation of the (Ga,In)As CBO. 
At \SI{100}{K} the difference of the (Ga,In)As CBO is $\approx -11\, \%$ while at \SI{290}{K} it is $\approx 11\, \%$ relative to the CBO at \SI{250}{K}. It is important to note, 
that this cannot be realized solely by a respective variation of the valence band offset (VBO) of Ga(As,Sb) without loosing the \mbox{type-I} character of the Ga(As,Sb) peak. It is known from lifetime measurements that the Ga(As,Sb) 
peak corresponds to a \mbox{type-I} transition at all temperatures \cite{dynamik}.

%Peakhoehen

 %alle PL ergebnisse von 26662-V3

%Concerning the relative PL peak intensities,
 Fig.~\ref{fig:Peakheights} shows a comparison between the experimental (blue) PL curves measured at \SI{250}{K} for two samples with 
d=\SI{4.8}{nm} (right) and d=\SI{3.5}{nm} (left) and the theoretical calculations (red). Minor deviations of the energy positions for the sample with d=\SI{3.5}{nm} are most probably due to slightly different concentrations.
However it is obvious, calculating the PL of the samples using equilibrium conditions for the carrier distributions does not well reproduce the relative heights of the peaks as can be seen in the upper part of Fig. \ref{fig:Peakheights}. The theoretically determined CT-PL peak is higher for both samples than the experimentally observed CT-peaks.    
These differences are indicative of the fact that the carrier system is not yet in complete quasi-equilibrium within the entire heterostructure. 
To account for this situation in the calculations and in order to match the heights of the experimental PL peaks, a weak non-equilibrium situation is assumed. In this case, the carrier distribution is described by a Fermi-Dirac distribution but the respective electronic temperature is higher than the temperature of the lattice.
The results are shown in the lower part of Fig.~\ref{fig:Peakheights}. We found a carrier temperature of \SI{380}{K} for the sample with d=\SI{4.8}{nm}, whereas for the structure with d=\SI{3.5}{nm} a carrier temperature of \SI{265}{K} is sufficient. The much higher effective carrier temperature in the sample with d=\SI{4.8}{nm} is most likely due to the less efficient tunneling processes between the wells. After excitation above the conduction band edge of GaAs, electrons and holes relax into the Ga(As,Sb) and (Ga,In)As quantum wells. However, since the tunneling probability decreases exponentially with increasing interlayer thickness, the relaxation into the equilibrium is less likely for the sample with d=\SI{4.8}{nm} than for the sample with a thinner barrier. 
Consequently, the carriers in the d=\SI{4.8}{nm} sample are on average at higher energies, i.e. they are effectively hotter during the emission process compared to those in the d=\SI{3.5}{nm} sample.\\

 \begin{figure}[!h]

 \includegraphics[width=8.5cm]{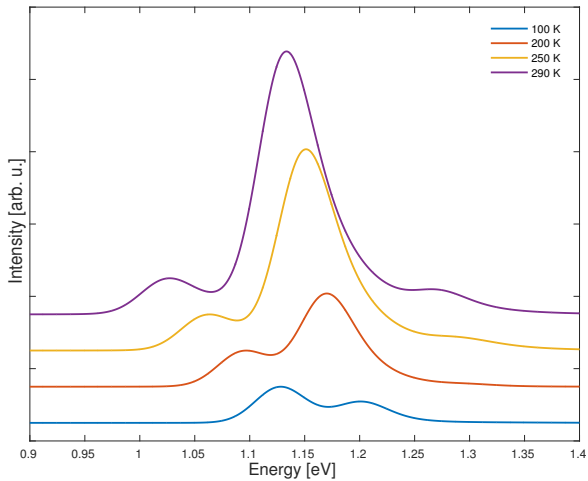}

 \caption{Theoretically obtained temperature-dependent PL spectra of the sample with d=\SI{4.8}{nm}.
	  The spectra are normalized to the \mbox{type-I}I PL and shifted vertically for visualization.}

 \label{fig:theo_PL26662}

\end{figure}

Since the PL is based on a non-dynamically calculation, it is assumed that all tunneling events which belong to the relaxation  
took place already. This causes that the (Ga,In)As transition does not appear in the theoretical PL at low temperatures but in the experimental spectrum.  
This can be seen by comparison of the PL spectrum of the sample with d = 4.8 nm in Fig.~\ref{fig:theo_PL26662} and Fig.~\ref{fig:fig3} at \SI{100}{K}. 
Contrary to this, a little (Ga,In)As transition appears in the theoretical PL at \SI{290}{K} as shoulder at the high energy side. 
This shoulder could not be observed in the experiment as it vanishes in background noise (see Fig.~\ref{fig:fig3}).\\

\begin{figure}

 \includegraphics[width=8.5cm]{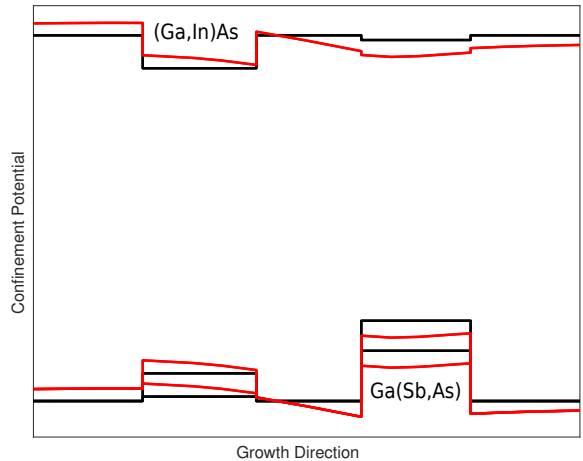}

 \caption{Density dependent confinement potential of the sample with d=\SI{4.8}{nm}. Depicted are the band edges for conduction band and the 
  heavy-hole and light-hole bands. 
  The black lines belong to a carrier density of $0.001\cdot 10^{12}/cm^2$ while 
  the red lines belong to a carrier density of $3.00\cdot 10^{12}/cm^2$ }

 \label{fig:bandverb}

\end{figure}

As discussed in chapter II, the excitation power dependent measurements show that the \mbox{type-I}I emission of the PL undergoes a larger blueshift 
compared to the \mbox{type-I} Ga(As,Sb) emission. This is caused by the electron accumulation in the (Ga,In)As well and hole accumulation in the Ga(As,Sb) well yielding a respective band bending due to the Coulomb interaction. In Fig.~\ref{fig:bandverb} the calculated confinement potential is depicted for two typical but different carrier densities. At low carrier density ($0.001\cdot 10^{12}/cm^2$; black line) the initial potential is almost unchanged, whereas at high density ($3.00\cdot 10^{12}/cm^2$, red curve) the band bending is obvious.
This band bending causes the stronger blueshift of the \mbox{type-I}I transition energy compared to the shift of the \mbox{type-I} transition in the Ga(As,Sb) well, 
which is solely due to the band filling.\\

%%%%%%%%%%%%%%%%%%%%%%%%%%%%%%%%%%%%%%%%%%%%%%%%%%%%%%%%%%%%%%%%%%%%%%%%%%%%%%%%%%%%%%%%%%%%%%%%%%%%%%%%%%%%%%%%%%%%%%%%%%%%%%%%%%%%%%%%%%%%%%%%%%%%%

\section{Conclusion}

We have experimentally and theoretically investigated a series of (Ga,In)As/GaAs/Ga(As,Sb) multi quantum well heterostructures with internal barriers of varying thickness.
The PL spectroscopy revealed that the \mbox{type-I}I emission in these samples is very effective.
By analyzing the temperature-dependence of the PL spectra we could find an important interplay of tunneling of holes and electrons to their respective lowest energy states and thermal repopulation of the higher-energy electron states in the Ga(As,Sb).
This leads to a characteristic PL behavior, while the \mbox{type-I}I and the Ga(As,Sb) transitions could be observed at room-temperature, the latter one disappears below \SI{100}{K} because it loses its holes to the (Ga,In)As.
The hole tunneling from the (Ga,In)As to the Ga(As,Sb) is, however, much less efficient so that the (Ga,In)As emission can be experimentally observed at low enough temperatures for thick internal barriers.
The experimental PL spectra of the sample with d=\SI{4.8}{nm} are theoretically well reproduced by employing a temperature dependent relative CBO between the wells and an increased carrier temperature where, however, in all cases the Ga(As,Sb) conduction band edge is situated energetically below that of GaAs.
The temperature dependent offset results in a decreasing energetic separation of the (Ga,In)As conduction band edge and the Ga(As,Sb)  valence band edge with increasing temperature. 
This causes a much stronger redshift of the \mbox{type-I}I emission with increasing temperature compared to the typical semiconductor bandgap shrinkage.   
Contrary to this, the \mbox{type-I}I emission at room temperature exhibits a larger blueshift compared to the \mbox{type-I} Ga(As,Sb) emission with increasing excitation densities. 
This is explained in terms of population dependent energy shifts of the hole end electron states of the quantum well layers.\\

%%%%%%%%%%%%%%%%%%%%%%%%%%%%%%%%%%%%%%%%%%%%%%%%%%%%%%%%%%%%%%%%%%%%%%%%%%%%%%%%%%%%%%%%%%%%%%%%%%%%%%%%%%%%%%%%%%%%%%%%%%%%%%%%%%%%%%%%%%%%%%%%%%%%%

\section*{Acknowledgments}

The work is a project of the Son\-der\-for\-schungs\-be\-reich 1083 funded by the Deutsche Forschungsgemeinschaft (DFG).

S.G. gratefully acknowledges DFG support in the framework of the GRK 1782. 

The Arizona contribution is based upon work supported by the Air Force Office of Scientific Research under award number BRI FA9550-14-1-0062.

We thank Dr. O. V\"ansk\"a for helpful discussions.

%%%%%%%%%%%%%%%%%%%%%%%%%%%%%%%%%%%%%%%%%%%%%%%%%%%%%%%%%%%%%%%%%%%%%%%%%%%%%%%%%%%%%%%%%%%%%%%%%%%%%%%%%%%%%%%%%%%%%%%%%%%%%%%%%%%%%%%%%%%%%%%%%%%%%

\bibliographystyle{apsrev4-1}

\bibliography{GaInAs_GaAsSb_references}

%merlin.mbs apsrev4-1.bst 2010-07-25 4.21a (PWD, AO, DPC) hacked
%Control: key (0)
%Control: author (72) initials jnrlst
%Control: editor formatted (1) identically to author
%Control: production of article title (-1) disabled
%Control: page (0) single
%Control: year (1) truncated
%Control: production of eprint (0) enabled
\begin{thebibliography}{19}%
\makeatletter
\providecommand \@ifxundefined [1]{%
 \@ifx{#1\undefined}
}%
\providecommand \@ifnum [1]{%
 \ifnum #1\expandafter \@firstoftwo
 \else \expandafter \@secondoftwo
 \fi
}%
\providecommand \@ifx [1]{%
 \ifx #1\expandafter \@firstoftwo
 \else \expandafter \@secondoftwo
 \fi
}%
\providecommand \natexlab [1]{#1}%
\providecommand \enquote  [1]{``#1''}%
\providecommand \bibnamefont  [1]{#1}%
\providecommand \bibfnamefont [1]{#1}%
\providecommand \citenamefont [1]{#1}%
\providecommand \href@noop [0]{\@secondoftwo}%
\providecommand \href [0]{\begingroup \@sanitize@url \@href}%
\providecommand \@href[1]{\@@startlink{#1}\@@href}%
\providecommand \@@href[1]{\endgroup#1\@@endlink}%
\providecommand \@sanitize@url [0]{\catcode `\\12\catcode `\$12\catcode
  `\&12\catcode `\#12\catcode `\^12\catcode `\_12\catcode `\%12\relax}%
\providecommand \@@startlink[1]{}%
\providecommand \@@endlink[0]{}%
\providecommand \url  [0]{\begingroup\@sanitize@url \@url }%
\providecommand \@url [1]{\endgroup\@href {#1}{\urlprefix }}%
\providecommand \urlprefix  [0]{URL }%
\providecommand \Eprint [0]{\href }%
\providecommand \doibase [0]{http://dx.doi.org/}%
\providecommand \selectlanguage [0]{\@gobble}%
\providecommand \bibinfo  [0]{\@secondoftwo}%
\providecommand \bibfield  [0]{\@secondoftwo}%
\providecommand \translation [1]{[#1]}%
\providecommand \BibitemOpen [0]{}%
\providecommand \bibitemStop [0]{}%
\providecommand \bibitemNoStop [0]{.\EOS\space}%
\providecommand \EOS [0]{\spacefactor3000\relax}%
\providecommand \BibitemShut  [1]{\csname bibitem#1\endcsname}%
\let\auto@bib@innerbib\@empty
%</preamble>
\bibitem [{\citenamefont {Pan}\ \emph {et~al.}(2010)\citenamefont {Pan},
  \citenamefont {Lin},\ and\ \citenamefont {Lee}}]{Pan2010}%
  \BibitemOpen
  \bibfield  {author} {\bibinfo {author} {\bibfnamefont {C.~H.}\ \bibnamefont
  {Pan}}, \bibinfo {author} {\bibfnamefont {S.~D.}\ \bibnamefont {Lin}}, \ and\
  \bibinfo {author} {\bibfnamefont {C.~P.}\ \bibnamefont {Lee}},\ }\href@noop
  {} {\bibfield  {journal} {\bibinfo  {journal} {Journal of Applied Physics}\
  }\textbf {\bibinfo {volume} {108}},\ \bibinfo {eid} {103105} (\bibinfo {year}
  {2010})}\BibitemShut {NoStop}%
\bibitem [{\citenamefont {Chang}\ \emph {et~al.}(2014)\citenamefont {Chang},
  \citenamefont {Li}, \citenamefont {Pan}, \citenamefont {Lu}, \citenamefont
  {Lee},\ and\ \citenamefont {Lin}}]{Chang2014}%
  \BibitemOpen
  \bibfield  {author} {\bibinfo {author} {\bibfnamefont {C.-H.}\ \bibnamefont
  {Chang}}, \bibinfo {author} {\bibfnamefont {Z.-L.}\ \bibnamefont {Li}},
  \bibinfo {author} {\bibfnamefont {C.-H.}\ \bibnamefont {Pan}}, \bibinfo
  {author} {\bibfnamefont {H.-T.}\ \bibnamefont {Lu}}, \bibinfo {author}
  {\bibfnamefont {C.-P.}\ \bibnamefont {Lee}}, \ and\ \bibinfo {author}
  {\bibfnamefont {S.-D.}\ \bibnamefont {Lin}},\ }\href@noop {} {\bibfield
  {journal} {\bibinfo  {journal} {Journal of Applied Physics}\ }\textbf
  {\bibinfo {volume} {115}},\ \bibinfo {eid} {063104} (\bibinfo {year}
  {2014})}\BibitemShut {NoStop}%
\bibitem [{\citenamefont {Pan}\ and\ \citenamefont {Lee}(2013)}]{Pan2013}%
  \BibitemOpen
  \bibfield  {author} {\bibinfo {author} {\bibfnamefont {C.~H.}\ \bibnamefont
  {Pan}}\ and\ \bibinfo {author} {\bibfnamefont {C.~P.}\ \bibnamefont {Lee}},\
  }\href@noop {} {\bibfield  {journal} {\bibinfo  {journal} {Journal of Applied
  Physics}\ }\textbf {\bibinfo {volume} {113}},\ \bibinfo {eid} {043112}
  (\bibinfo {year} {2013})}\BibitemShut {NoStop}%
\bibitem [{\citenamefont {Ripalda}\ \emph {et~al.}(2005)\citenamefont
  {Ripalda}, \citenamefont {Granados}, \citenamefont {Gonz\'{a}lez},
  \citenamefont {S\'{a}nchez}, \citenamefont {Molina},\ and\ \citenamefont
  {Garc\'{i}a}}]{Ripalda2005}%
  \BibitemOpen
  \bibfield  {author} {\bibinfo {author} {\bibfnamefont {J.~M.}\ \bibnamefont
  {Ripalda}}, \bibinfo {author} {\bibfnamefont {D.}~\bibnamefont {Granados}},
  \bibinfo {author} {\bibfnamefont {Y.}~\bibnamefont {Gonz\'{a}lez}}, \bibinfo
  {author} {\bibfnamefont {A.~M.}\ \bibnamefont {S\'{a}nchez}}, \bibinfo
  {author} {\bibfnamefont {S.~I.}\ \bibnamefont {Molina}}, \ and\ \bibinfo
  {author} {\bibfnamefont {J.~M.}\ \bibnamefont {Garc\'{i}a}},\ }\href@noop {}
  {\bibfield  {journal} {\bibinfo  {journal} {Applied Physics Letters}\
  }\textbf {\bibinfo {volume} {87}},\ \bibinfo {eid} {202108} (\bibinfo {year}
  {2005})}\BibitemShut {NoStop}%
\bibitem [{\citenamefont {M\"{o}ller}\ \emph {et~al.}(2016)\citenamefont
  {M\"{o}ller}, \citenamefont {Fuchs}, \citenamefont {Berger}, \citenamefont
  {Ruiz~Perez}, \citenamefont {Koch}, \citenamefont {Hader}, \citenamefont
  {Moloney}, \citenamefont {Koch},\ and\ \citenamefont {Stolz}}]{Moeller2016}%
  \BibitemOpen
  \bibfield  {author} {\bibinfo {author} {\bibfnamefont {C.}~\bibnamefont
  {M\"{o}ller}}, \bibinfo {author} {\bibfnamefont {C.}~\bibnamefont {Fuchs}},
  \bibinfo {author} {\bibfnamefont {C.}~\bibnamefont {Berger}}, \bibinfo
  {author} {\bibfnamefont {A.}~\bibnamefont {Ruiz~Perez}}, \bibinfo {author}
  {\bibfnamefont {M.}~\bibnamefont {Koch}}, \bibinfo {author} {\bibfnamefont
  {J.}~\bibnamefont {Hader}}, \bibinfo {author} {\bibfnamefont {J.~V.}\
  \bibnamefont {Moloney}}, \bibinfo {author} {\bibfnamefont {S.~W.}\
  \bibnamefont {Koch}}, \ and\ \bibinfo {author} {\bibfnamefont
  {W.}~\bibnamefont {Stolz}},\ }\href@noop {} {\bibfield  {journal} {\bibinfo
  {journal} {Applied Physics Letters}\ }\textbf {\bibinfo {volume} {108}},\
  \bibinfo {eid} {071102} (\bibinfo {year} {2016})}\BibitemShut {NoStop}%
\bibitem [{\citenamefont {Berger}\ \emph {et~al.}(2015)\citenamefont {Berger},
  \citenamefont {M\"{o}ller}, \citenamefont {Hens}, \citenamefont {Fuchs},
  \citenamefont {Stolz}, \citenamefont {Koch}, \citenamefont {Ruiz~Perez},
  \citenamefont {Hader},\ and\ \citenamefont {Moloney}}]{Berger2015}%
  \BibitemOpen
  \bibfield  {author} {\bibinfo {author} {\bibfnamefont {C.}~\bibnamefont
  {Berger}}, \bibinfo {author} {\bibfnamefont {C.}~\bibnamefont {M\"{o}ller}},
  \bibinfo {author} {\bibfnamefont {P.}~\bibnamefont {Hens}}, \bibinfo {author}
  {\bibfnamefont {C.}~\bibnamefont {Fuchs}}, \bibinfo {author} {\bibfnamefont
  {W.}~\bibnamefont {Stolz}}, \bibinfo {author} {\bibfnamefont {S.~W.}\
  \bibnamefont {Koch}}, \bibinfo {author} {\bibfnamefont {A.}~\bibnamefont
  {Ruiz~Perez}}, \bibinfo {author} {\bibfnamefont {J.}~\bibnamefont {Hader}}, \
  and\ \bibinfo {author} {\bibfnamefont {J.~V.}\ \bibnamefont {Moloney}},\
  }\href@noop {} {\bibfield  {journal} {\bibinfo  {journal} {AIP Advances}\
  }\textbf {\bibinfo {volume} {5}},\ \bibinfo {eid} {047105} (\bibinfo {year}
  {2015})}\BibitemShut {NoStop}%
\bibitem [{\citenamefont {Zegrya}\ and\ \citenamefont
  {Andreev}(1995)}]{Zegrya1995}%
  \BibitemOpen
  \bibfield  {author} {\bibinfo {author} {\bibfnamefont {G.~G.}\ \bibnamefont
  {Zegrya}}\ and\ \bibinfo {author} {\bibfnamefont {A.~D.}\ \bibnamefont
  {Andreev}},\ }\href@noop {} {\bibfield  {journal} {\bibinfo  {journal}
  {Applied Physics Letters}\ }\textbf {\bibinfo {volume} {67}},\ \bibinfo
  {pages} {2681} (\bibinfo {year} {1995})}\BibitemShut {NoStop}%
\bibitem [{\citenamefont {Gies}\ \emph {et~al.}(2015)\citenamefont {Gies},
  \citenamefont {Kruska}, \citenamefont {Berger}, \citenamefont {Hens},
  \citenamefont {Fuchs}, \citenamefont {Ruiz~Perez}, \citenamefont {Rosemann},
  \citenamefont {Veletas}, \citenamefont {Chatterjee}, \citenamefont {Stolz},
  \citenamefont {Koch}, \citenamefont {Hader}, \citenamefont {Moloney},\ and\
  \citenamefont {Heimbrodt}}]{Gies2015}%
  \BibitemOpen
  \bibfield  {author} {\bibinfo {author} {\bibfnamefont {S.}~\bibnamefont
  {Gies}}, \bibinfo {author} {\bibfnamefont {C.}~\bibnamefont {Kruska}},
  \bibinfo {author} {\bibfnamefont {C.}~\bibnamefont {Berger}}, \bibinfo
  {author} {\bibfnamefont {P.}~\bibnamefont {Hens}}, \bibinfo {author}
  {\bibfnamefont {C.}~\bibnamefont {Fuchs}}, \bibinfo {author} {\bibfnamefont
  {A.}~\bibnamefont {Ruiz~Perez}}, \bibinfo {author} {\bibfnamefont {N.~W.}\
  \bibnamefont {Rosemann}}, \bibinfo {author} {\bibfnamefont {J.}~\bibnamefont
  {Veletas}}, \bibinfo {author} {\bibfnamefont {S.}~\bibnamefont {Chatterjee}},
  \bibinfo {author} {\bibfnamefont {W.}~\bibnamefont {Stolz}}, \bibinfo
  {author} {\bibfnamefont {S.~W.}\ \bibnamefont {Koch}}, \bibinfo {author}
  {\bibfnamefont {J.}~\bibnamefont {Hader}}, \bibinfo {author} {\bibfnamefont
  {J.~V.}\ \bibnamefont {Moloney}}, \ and\ \bibinfo {author} {\bibfnamefont
  {W.}~\bibnamefont {Heimbrodt}},\ }\href@noop {} {\bibfield  {journal}
  {\bibinfo  {journal} {Applied Physics Letters}\ }\textbf {\bibinfo {volume}
  {107}},\ \bibinfo {eid} {182104} (\bibinfo {year} {2015})}\BibitemShut
  {NoStop}%
\bibitem [{\citenamefont {Springer}\ \emph {et~al.}(2016)\citenamefont
  {Springer}, \citenamefont {Gies}, \citenamefont {Hens}, \citenamefont
  {Fuchs}, \citenamefont {Han}, \citenamefont {Hader}, \citenamefont {Moloney},
  \citenamefont {Stolz}, \citenamefont {Volz}, \citenamefont {Koch},\ and\
  \citenamefont {Heimbrodt}}]{Springer2016}%
  \BibitemOpen
  \bibfield  {author} {\bibinfo {author} {\bibfnamefont {P.}~\bibnamefont
  {Springer}}, \bibinfo {author} {\bibfnamefont {S.}~\bibnamefont {Gies}},
  \bibinfo {author} {\bibfnamefont {P.}~\bibnamefont {Hens}}, \bibinfo {author}
  {\bibfnamefont {C.}~\bibnamefont {Fuchs}}, \bibinfo {author} {\bibfnamefont
  {H.}~\bibnamefont {Han}}, \bibinfo {author} {\bibfnamefont {J.}~\bibnamefont
  {Hader}}, \bibinfo {author} {\bibfnamefont {J.}~\bibnamefont {Moloney}},
  \bibinfo {author} {\bibfnamefont {W.}~\bibnamefont {Stolz}}, \bibinfo
  {author} {\bibfnamefont {K.}~\bibnamefont {Volz}}, \bibinfo {author}
  {\bibfnamefont {S.}~\bibnamefont {Koch}}, \ and\ \bibinfo {author}
  {\bibfnamefont {W.}~\bibnamefont {Heimbrodt}},\ }\href@noop {} {\bibfield
  {journal} {\bibinfo  {journal} {Journal of Luminescence}\ }\textbf {\bibinfo
  {volume} {175}},\ \bibinfo {pages} {255 } (\bibinfo {year}
  {2016})}\BibitemShut {NoStop}%
\bibitem [{\citenamefont {Antypas}\ and\ \citenamefont
  {James}(1970)}]{Antypas1970}%
  \BibitemOpen
  \bibfield  {author} {\bibinfo {author} {\bibfnamefont {G.~A.}\ \bibnamefont
  {Antypas}}\ and\ \bibinfo {author} {\bibfnamefont {L.~W.}\ \bibnamefont
  {James}},\ }\href@noop {} {\bibfield  {journal} {\bibinfo  {journal} {Journal
  of Applied Physics}\ }\textbf {\bibinfo {volume} {41}},\ \bibinfo {pages}
  {2165} (\bibinfo {year} {1970})}\BibitemShut {NoStop}%
\bibitem [{\citenamefont {Nahory}\ \emph {et~al.}(1977)\citenamefont {Nahory},
  \citenamefont {Pollack}, \citenamefont {DeWinter},\ and\ \citenamefont
  {Williams}}]{Nahory1977}%
  \BibitemOpen
  \bibfield  {author} {\bibinfo {author} {\bibfnamefont {R.~E.}\ \bibnamefont
  {Nahory}}, \bibinfo {author} {\bibfnamefont {M.~A.}\ \bibnamefont {Pollack}},
  \bibinfo {author} {\bibfnamefont {J.~C.}\ \bibnamefont {DeWinter}}, \ and\
  \bibinfo {author} {\bibfnamefont {K.~M.}\ \bibnamefont {Williams}},\
  }\href@noop {} {\bibfield  {journal} {\bibinfo  {journal} {Journal of Applied
  Physics}\ }\textbf {\bibinfo {volume} {48}},\ \bibinfo {pages} {1607}
  (\bibinfo {year} {1977})}\BibitemShut {NoStop}%
\bibitem [{\citenamefont {Goetz}\ \emph {et~al.}(1983)\citenamefont {Goetz},
  \citenamefont {Bimberg}, \citenamefont {J\"{u}rgensen}, \citenamefont
  {Selders}, \citenamefont {Solomonov}, \citenamefont {Glinskii},\ and\
  \citenamefont {Razeghi}}]{Goetz1983}%
  \BibitemOpen
  \bibfield  {author} {\bibinfo {author} {\bibfnamefont {K.-H.}\ \bibnamefont
  {Goetz}}, \bibinfo {author} {\bibfnamefont {D.}~\bibnamefont {Bimberg}},
  \bibinfo {author} {\bibfnamefont {H.}~\bibnamefont {J\"{u}rgensen}}, \bibinfo
  {author} {\bibfnamefont {J.}~\bibnamefont {Selders}}, \bibinfo {author}
  {\bibfnamefont {A.~V.}\ \bibnamefont {Solomonov}}, \bibinfo {author}
  {\bibfnamefont {G.~F.}\ \bibnamefont {Glinskii}}, \ and\ \bibinfo {author}
  {\bibfnamefont {M.}~\bibnamefont {Razeghi}},\ }\href@noop {} {\bibfield
  {journal} {\bibinfo  {journal} {Journal of Applied Physics}\ }\textbf
  {\bibinfo {volume} {54}},\ \bibinfo {pages} {4543} (\bibinfo {year}
  {1983})}\BibitemShut {NoStop}%
\bibitem [{\citenamefont {Tatebayashi}\ \emph {et~al.}(2008)\citenamefont
  {Tatebayashi}, \citenamefont {Liang}, \citenamefont {Laghumavarapu},
  \citenamefont {Bussian}, \citenamefont {Htoon}, \citenamefont {Klimov},
  \citenamefont {Balakrishnan}, \citenamefont {Dawson},\ and\ \citenamefont
  {Huffaker}}]{Tatebayashi2008}%
  \BibitemOpen
  \bibfield  {author} {\bibinfo {author} {\bibfnamefont {J.}~\bibnamefont
  {Tatebayashi}}, \bibinfo {author} {\bibfnamefont {B.~L.}\ \bibnamefont
  {Liang}}, \bibinfo {author} {\bibfnamefont {R.~B.}\ \bibnamefont
  {Laghumavarapu}}, \bibinfo {author} {\bibfnamefont {D.~A.}\ \bibnamefont
  {Bussian}}, \bibinfo {author} {\bibfnamefont {H.}~\bibnamefont {Htoon}},
  \bibinfo {author} {\bibfnamefont {V.}~\bibnamefont {Klimov}}, \bibinfo
  {author} {\bibfnamefont {G.}~\bibnamefont {Balakrishnan}}, \bibinfo {author}
  {\bibfnamefont {L.~R.}\ \bibnamefont {Dawson}}, \ and\ \bibinfo {author}
  {\bibfnamefont {D.~L.}\ \bibnamefont {Huffaker}},\ }\href@noop {} {\bibfield
  {journal} {\bibinfo  {journal} {Nanotechnology}\ }\textbf {\bibinfo {volume}
  {19}},\ \bibinfo {pages} {295704} (\bibinfo {year} {2008})}\BibitemShut
  {NoStop}%
\bibitem [{\citenamefont {Kira}\ \emph {et~al.}(1999)\citenamefont {Kira},
  \citenamefont {Jahnke}, \citenamefont {Hoyer},\ and\ \citenamefont
  {Koch}}]{Kira1999189}%
  \BibitemOpen
  \bibfield  {author} {\bibinfo {author} {\bibfnamefont {M.}~\bibnamefont
  {Kira}}, \bibinfo {author} {\bibfnamefont {F.}~\bibnamefont {Jahnke}},
  \bibinfo {author} {\bibfnamefont {W.}~\bibnamefont {Hoyer}}, \ and\ \bibinfo
  {author} {\bibfnamefont {S.}~\bibnamefont {Koch}},\ }\href {\doibase
  http://dx.doi.org/10.1016/S0079-6727(99)00008-7} {\bibfield  {journal}
  {\bibinfo  {journal} {Progress in Quantum Electronics}\ }\textbf {\bibinfo
  {volume} {23}},\ \bibinfo {pages} {189 } (\bibinfo {year}
  {1999})}\BibitemShut {NoStop}%
\bibitem [{\citenamefont {Hader}\ \emph {et~al.}(1997)\citenamefont {Hader},
  \citenamefont {Linder},\ and\ \citenamefont {D\"ohler}}]{PhysRevB.55.6960}%
  \BibitemOpen
  \bibfield  {author} {\bibinfo {author} {\bibfnamefont {J.}~\bibnamefont
  {Hader}}, \bibinfo {author} {\bibfnamefont {N.}~\bibnamefont {Linder}}, \
  and\ \bibinfo {author} {\bibfnamefont {G.~H.}\ \bibnamefont {D\"ohler}},\
  }\href {\doibase 10.1103/PhysRevB.55.6960} {\bibfield  {journal} {\bibinfo
  {journal} {Phys. Rev. B}\ }\textbf {\bibinfo {volume} {55}},\ \bibinfo
  {pages} {6960} (\bibinfo {year} {1997})}\BibitemShut {NoStop}%
\bibitem [{\citenamefont {Chow}\ and\ \citenamefont {Koch}(1999)}]{ChowKoch}%
  \BibitemOpen
  \bibfield  {author} {\bibinfo {author} {\bibfnamefont {W.~W.}\ \bibnamefont
  {Chow}}\ and\ \bibinfo {author} {\bibfnamefont {S.~W.}\ \bibnamefont
  {Koch}},\ }\href@noop {} {\emph {\bibinfo {title} {Semiconductor-Laser
  Fundamentals: Physics of the Gain Materials}}}\ (\bibinfo  {publisher}
  {Springer},\ \bibinfo {address} {Berlin, Heidelberg, New York},\ \bibinfo
  {year} {1999})\BibitemShut {NoStop}%
\bibitem [{\citenamefont {Hader}\ \emph {et~al.}(2003)\citenamefont {Hader},
  \citenamefont {Koch},\ and\ \citenamefont {Moloney}}]{Hader2003513}%
  \BibitemOpen
  \bibfield  {author} {\bibinfo {author} {\bibfnamefont {J.}~\bibnamefont
  {Hader}}, \bibinfo {author} {\bibfnamefont {S.}~\bibnamefont {Koch}}, \ and\
  \bibinfo {author} {\bibfnamefont {J.}~\bibnamefont {Moloney}},\ }\href
  {\doibase http://dx.doi.org/10.1016/S0038-1101(02)00405-7} {\bibfield
  {journal} {\bibinfo  {journal} {Solid-State Electronics}\ }\textbf {\bibinfo
  {volume} {47}},\ \bibinfo {pages} {513 } (\bibinfo {year}
  {2003})}\BibitemShut {NoStop}%
\bibitem [{\citenamefont {Vurgaftman}\ \emph {et~al.}(2001)\citenamefont
  {Vurgaftman}, \citenamefont {Meyer},\ and\ \citenamefont
  {Ram-Mohan}}]{vurg2001}%
  \BibitemOpen
  \bibfield  {author} {\bibinfo {author} {\bibfnamefont {I.}~\bibnamefont
  {Vurgaftman}}, \bibinfo {author} {\bibfnamefont {J.~R.}\ \bibnamefont
  {Meyer}}, \ and\ \bibinfo {author} {\bibfnamefont {L.~R.}\ \bibnamefont
  {Ram-Mohan}},\ }\href {\doibase http://dx.doi.org/10.1063/1.1368156}
  {\bibfield  {journal} {\bibinfo  {journal} {Journal of Applied Physics}\
  }\textbf {\bibinfo {volume} {89}},\ \bibinfo {pages} {5815} (\bibinfo {year}
  {2001})}\BibitemShut {NoStop}%
\bibitem [{\citenamefont {{Gies}}\ \emph {et~al.}(2016)\citenamefont {{Gies}},
  \citenamefont {{Holz}}, \citenamefont {{Fuchs}}, \citenamefont {{Stolz}},\
  and\ \citenamefont {{Heimbrodt}}}]{dynamik}%
  \BibitemOpen
  \bibfield  {author} {\bibinfo {author} {\bibfnamefont {S.}~\bibnamefont
  {{Gies}}}, \bibinfo {author} {\bibfnamefont {B.}~\bibnamefont {{Holz}}},
  \bibinfo {author} {\bibfnamefont {C.}~\bibnamefont {{Fuchs}}}, \bibinfo
  {author} {\bibfnamefont {W.}~\bibnamefont {{Stolz}}}, \ and\ \bibinfo
  {author} {\bibfnamefont {W.}~\bibnamefont {{Heimbrodt}}},\ }\href@noop {}
  {\bibfield  {journal} {\bibinfo  {journal} {ArXiv e-prints}\ } (\bibinfo
  {year} {2016})},\ \Eprint {http://arxiv.org/abs/1609.09255} {arXiv:1609.09255
  [cond-mat.mes-hall]} \BibitemShut {NoStop}%
\end{thebibliography}%

%%%%%%%%%%%%%%%%%%%%%%%%%%%%%%%%%%%%%%%%%%%%%%%%%%

\end{document}